# Analysis of the temperature dependence of the electron spin resonance linewidth in exchange-coupled magnetic insulators

M. Acikgoz [a] and D. L. Huber [b]

[a] Department of Chemistry, Rutgers University-Newark, Newark, NJ 07102, USA. E-mail address: muhammed.acikgoz@rutgers.edu

[b] Department of Physics, University of Wisconsin-Madison, Madison, WI 53706, USA. E-mail address: dhuber@wisc.edu

**Abstract**

We analyze the temperature dependence of the electron spin resonance linewidth in exchange-coupled magnetic insulators using results from $Co_3O_4$ as an example. The focus is on separating the contributions from spin-spin interactions, spin-one-phonon interactions and spin-two-phonon interactions. Expressing the linewidth as a sum of the three contributions, varying as const., $BT$, and $CT^2$, respectively, we use a least-squares fit over the temperature range $50\ K \leq T \leq 500\ K$ to obtain values of the three components. It is found that the spin-spin mechanism is dominant below $100\ K$, while the two-phonon mechanism is most important above $250\ K$. In the intermediate region, all three mechanisms make significant contributions. The success of the high temperature approximations for the one and two-phonon terms, which occurs well below the Debye temperature of $525\ K$ in $Co_3O_4$, is attributed to extreme exchange narrowing of the bandwidth of phonons contributing to the linewidth.

The authors declare that there is no conflict of interest regarding publication of this paper.

## 1. Introduction

Studies of the electron spin resonance (ESR) linewidth in the magnetic insulator $CrBr_3$ revealed a liner temperature dependence well above the critical region [1]. Subsequent analysis show that the linear temperature dependence was associated with the spin-phonon interaction [2,3]. Since $Cr^{3+}$ is not an S-state ion, the interactions between the spins and phonons are comparatively strong in contrast to S-state systems such as $MnF_2$ where the linewidth approaches a constant value at high temperatures. Recently, results have been reported which indicate that in $Co_3O_4$, the linewidth begins to vary quadratically with increasing temperature [4,5] suggesting the growing importance of two-phonon processes. In this note we demonstrate how one can separate the contributions to the linewidth from the three processes.

## 2. Analysis

As pointed out in [3], above the critical region, the ESR linewidth in magnetic insulators can be expressed in the form

$$\Delta H(T) = (\chi_0(T)/\chi(T))[A + BT + CT^2] \tag{1}$$

where $\chi_0(T)$ denotes the Curie susceptibility and $\chi(T)$ is the static susceptibility. The letters $A$, $B$ and $C$ refer to the spin-spin, spin-one-phonon, and spin-two-phonon contributions, respectively. In our analysis of $Co_3O_4$ we can use the Curie-Weiss approximation for the susceptibility so that the first factor in (1) takes the form

$$\chi_0(T)/\chi(T) = (T-\theta)/T \tag{2}$$

We analyze the data for $Co_3O_4$ from [5] over the range 50 K $\leq T \leq$ 500 K with $\theta = -110\ K$ [5]. We assume that $A$, $B$ and $C$ are temperature-independent above 50 K., a high-temperature approximation consistent with the Néel temperature, 39 K. In Fig. 1 we show the results obtained from fitting the X-band linewidth data shown in Fig. 4 of [5]. The corresponding fitting parameters are

$$A = 3.607 \times 10^2\ Oe \tag{3}$$

$$B = 1.225\ OeK^{-1} \tag{4}$$

$$C = 8.570 \times 10^{-3}\ OeK^{-2} \tag{5}$$

We discuss these results in the following section.

## 3. Discussion

As noted, the temperature dependence of the ESR linewidth of $Co_3O_4$ above the critical region reflects the interplay of spin-spin and spin-lattice interactions. In Fig. 2, we plot the temperature dependence of the three terms contributing to the product $[T/(T+110)]\Delta H(T)$. It is apparent that below 100 K, the linewidth is dominated by the contribution from spin-spin interactions, while the two-phonon processes are dominant above 250 K [4,5]. The range 100 K

< $T$ < 250 $K$ is a cross-over region where all three processes are making a significant contribution to the width.

It should be noted that the functional forms $BT$ and $CT^2$ are high-temperature approximations to the expressions for the one-phonon and two-phonon processes. The explanation for the high temperature form is discussed in [3] where it is shown that exchange interactions between the magnetic ions limit the frequencies of the phonons contributing to the linewidth in a one-phonon process. Similar arguments apply to two-phonon processes as well. An estimate of the cut-off energy of the phonons in a system with nearest-neighbor exchange interactions is given in [3] and takes the form

$$E_{phonon}^{cut-off} = (2S(S+1)n_{nn}J_{nn}^2)^{1/2} \qquad (6)$$

where $J_{nn}$ is the nearest-neighbor exchange interaction and $n_{nn}$ is the number of nearest neighbors. With $S = 3/2$, $n_{nn} = 4$ and $J_{nn} = 11.7\ K$ [4], we obtain a cut-off temperature equal to 64 $K$. As a result, we expect the high temperature form for the spin-phonon contributions is appropriate for $T > 60 – 70\ K$. For comparison, we note that the Debye temperature is the nominal boundary of the high temperature regime for acoustic phonons. In the case of Co3O4 the experimental Debye temperature is 525 $K$ [6], eight times larger than the exchange cut-off, indicating an extreme exchange narrowing of the width of the phonon band contributing to the linewidth.

### 4. Summary

We have outlined an approach for separating the contributions to the high-temperature ESR linewidth in magnetic insulators that are associated with spin-spin and one- and two-phonon processes. The analysis is applicable above the critical regime, which we loosely identify as the region where the Curie-Weiss approximation is useful, provided the temperature exceeds a cutoff temperature of the band of contributing phonons. ESR studies of Co3O4 show evidence of an extreme exchange narrowing of the spectrum of contributing phonons.

**Acknowledgment**

The authors would like to thank Z. Seidov for the ESR measurements.

**Figure captions**

Fig. 1. $Co_3O_4$. ESR linewidth vs $T$. The data points are from [5]. The solid curve is a three-parameter least squares fit described in the text.

Fig. 2. $Co_3O_4$. Contributions to $[T/(T+110)]\Delta H(T)$ from spin-spin interactions, $A$; one-phonon processes, $BT$; two-phonon processes, $CT^2$

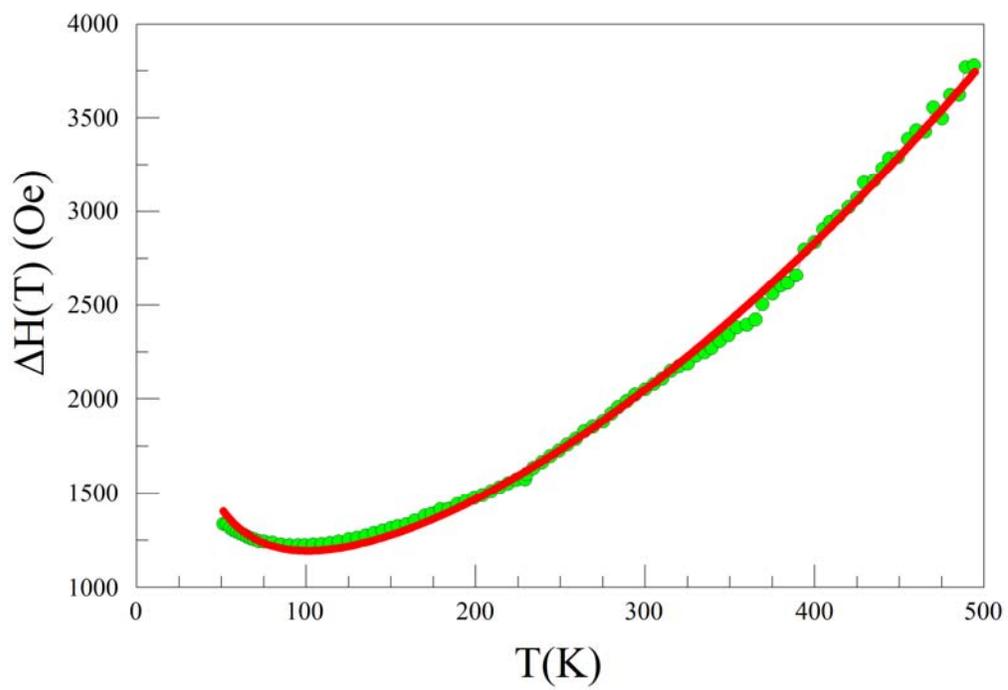

**Fig. 1**

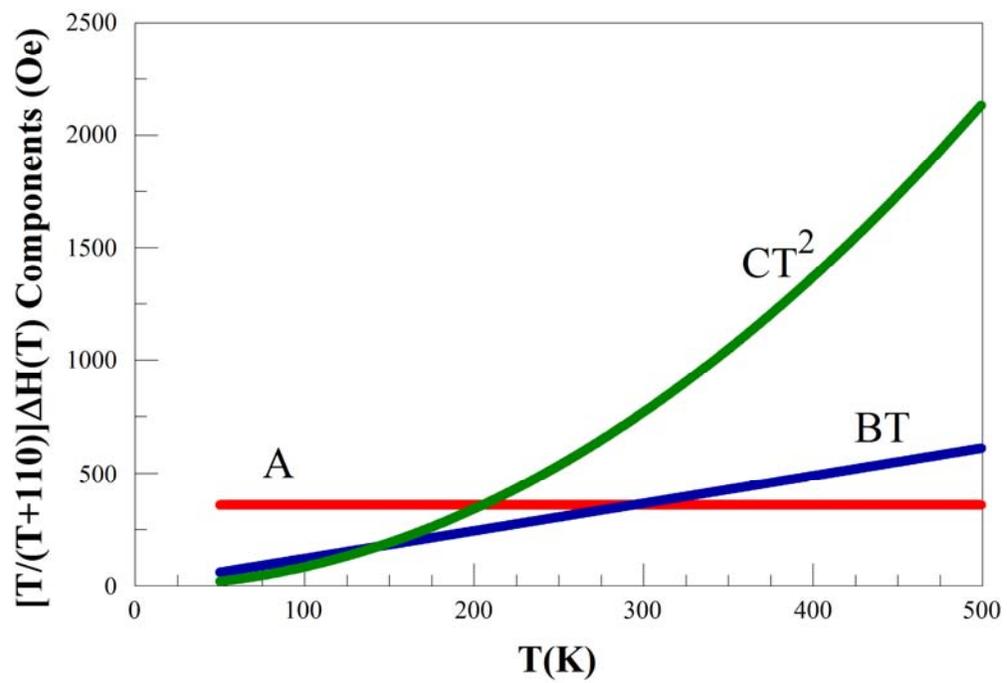

**Fig. 2**